\documentclass[epj,final]{svjour}
\smartqed  
\RequirePackage{graphicx}
\begin{document}

\title{Comparison of Tsallis statistics with the Tsallis-factorized statistics in the ultrarelativistic $pp$ collisions}

\author{A.S.~Parvan}

\institute{Bogoliubov Laboratory of Theoretical Physics, Joint Institute for Nuclear Research, Dubna, Russian Federation \and
Department of Theoretical Physics, Horia Hulubei National Institute of Physics and Nuclear Engineering, Bucharest-Magurele, Romania}

\date{Received: date / Revised version: date}

\abstract{The Tsallis statistics was applied to describe the experimental data on the transverse momentum distributions of hadrons. We considered the energy dependence of the parameters of the Tsallis-factorized statistics, which is now widely used for the description of the experimental transverse momentum distributions of hadrons, and the Tsallis statistics for the charged pions produced in $pp$ collisions at high energies. We found that the results of the Tsallis-factorized statistics deviate from the results of the Tsallis statistics only at low NA61/SHINE energies when the value of the entropic parameter is close to unity. At higher energies, when the value of the entropic parameter deviates essentially from unity, the Tsallis-factorized statistics satisfactorily recovers the results of the Tsallis statistics.
} 
\titlerunning{Comparison of Tsallis statistics with the Tsallis-factorized statistics}
\authorrunning{A.S.~Parvan}
\maketitle
\section{Introduction}\label{sec1}
There is currently great interest in analyzing the experimental data on the transverse momentum distributions of hadrons created in the proton-proton and heavy-ion collisions at high energies by different approximations of the Tsallis statistics~\cite{Alice1a,Alice1b,Cms2012,Cms2014,Cleymans12a,Cleymans2012,Cleymans13,Rybczynski14,Parvan14,Parvan16a,Liu2014,Wibig2014,Azmi2015,McLerran2016,Li2013,Tang09}. The momentum distribution of the Tsallis-factorized statistics given in~\cite{Cleymans12a,Cleymans2012} is very useful for the description of the experimental data because of its simple analytical formula. Although the momentum distributions of the Tsallis-factorized statistics were successfully used in analyzing the experimental data, the Tsallis statistics itself~\cite{Tsal88,Tsal98} has not been applied to the treatment of the transverse momentum distributions of hadrons yet (see Ref.~\cite{Parvan16b} and references therein).

The Tsallis statistics~\cite{Tsal88,Tsal98} is defined on the basis of the constrained extrema of the Tsallis entropy generalized from the Boltzmann-Gibbs entropy with respect to the many-body distribution functions. However, the Tsallis-factorized statistics~\cite{Cleymans12a,Cleymans2012} is defined on the basis of the constrained extrema of the Tsallis entropy of the ideal gas generalized from the Boltzmann-Gibbs entropy of the ideal gas with respect to the single-particle distribution functions. These two statistics give the different results for the thermodynamic quantities. It should be stressed that these two different extremizations give the same results for the mean occupation numbers only in the Boltzmann-Gibbs statistics. But, for the Tsallis statistics this property of the ideal gas of the Boltzmann-Gibbs statistics is not preserved. Thus, the Tsallis-factorized statistics and the Tsallis statistics are not equivalent. They are different~\cite{Parvan16b}. The Tsallis statistics~\cite{Tsal88} is thermodynamically self-consistent in the thermodynamic limit. This was demonstrated in the framework of the different statistical ensembles in Refs.~\cite{Parvan06a,Parvan06b,Parvan2015a,Parvan2015}. The main theoretical difficulty of the Tsallis-factorized statistics is that the Tsallis-factorized statistics is defined only for a particular case of the ideal gas and that for each statistics of particles (Maxwell - Boltzmann, Bose-Einstein and Fermi-Dirac) the entropy of the Tsallis-factorized statistics is redefined.

The analytical expression for the ultrarelativistic transverse momentum distribution of the Tsallis statistics was obtained in~\cite{Parvan16b}. It was demonstrated that the transverse momentum distribution of the Tsallis-factorized statistics~\cite{Cleymans12a,Cleymans2012}, which is used for the description of the experimental transverse momentum spectra of hadrons at high energies, in the ultrarelativistic case is not equivalent to the transverse momentum distributions of the Tsallis and the Tsallis-$2$ statistics. This distribution is similar only to the transverse momentum distribution of the Tsallis-$2$ statistics in the zeroth term approximation and to the transverse momentum distribution of the Tsallis statistics in the zeroth term approximation with transformation of the parameter $q$ to the parameter $1/q_{c}$.

The main purpose of this study is to apply the analytical expression for the ultrarelativistic transverse momentum distribution of the Tsallis statistics given in Ref.~\cite{Parvan16b} to describe the experimental data for the charged pions measured in $pp$ collisions at LHC and RHIC energies and to compare numerically the parameters of the Tsallis statistics with the parameters of the Tsallis-factorized statistics~\cite{Cleymans12a,Cleymans2012} at different energies of $pp$ collision.

The structure of the paper is as follows. In Section~\ref{sec2}, we define the formulas and analyze the data and the results. The main conclusions are given in the final section.

\section{Transverse momentum spectrum}\label{sec2}
Let us compare numerically the transverse momentum distributions of the Tsallis statistics with the transverse momentum distributions of the Tsallis-factorized statistics and apply them to describe the experimental data for the charged pions produced in $pp$ collisions with energy in the range from $\sqrt{s}=6.3$ GeV to $7000$ GeV. In this paper, we consider only the ultrarelativistic case. The single-particle distribution function of the transverse momentum $p_{T}$ and rapidity $y$ for the Maxwell-Boltzmann ultrarelativistic ideal gas of the Tsallis statistics in the grand canonical ensemble can be written as (see Ref.~\cite{Parvan16b})
\begin{eqnarray}\label{1}
  &&\frac{d^{2}N}{dp_{T}dy} = \frac{gV}{(2\pi)^{2}} p_{T}^{2} \cosh y  \   \sum\limits_{N=0}^{N_{0}} \frac{\tilde{\omega}^{N}}{N!} h_{0}(0) \nonumber \\
   &&\;\;\;\; \left[1+\frac{q-1}{q}\frac{\Lambda-p_{T} \cosh y +\mu (N+1)}{T}\right]^{\frac{1}{q-1}+3N},
\end{eqnarray}
where
\begin{equation}\label{2}
   h_{0}(0) = \frac{\left(\frac{q}{1-q}\right)^{3N}  \Gamma\left(\frac{1}{1-q}-3N\right)}{\Gamma\left(\frac{1}{1-q}\right)}, \qquad  q<1
\end{equation}
and
\begin{equation}\label{3}
   h_{0}(0) =\frac{\left(\frac{q}{q-1}\right)^{3N}  \Gamma\left(\frac{q}{q-1}\right)}{\Gamma\left(\frac{q}{q-1}+3N\right)},   \qquad  q>1.
\end{equation}
Here, $\tilde{\omega}=gVT^{3}/\pi^{2}$, $\mu, T$ and $V$ are the chemical potential, the temperature and the volume, respectively, $g$ is the spin degeneracy factor and $q\in\mathbf{R}$ is a real parameter taking values $0<q<\infty$. The norm function $\Lambda$ is determined from the norm equation
\begin{equation}\label{4}
   \sum\limits_{N=0}^{N_{0}} \frac{\tilde{\omega}^{N}}{N!} h_{0}(0) \left[1+\frac{q-1}{q}\frac{\Lambda+\mu N}{T}\right]^{\frac{1}{q-1}+3N}=1,
\end{equation}
where the upper bound of summation $N_{0}$ for $q<1$ follows from the inequality $N<1/(3(1-q))$ and the inflection point of the logarithm of the function
\begin{equation}\label{5}
  \phi(N) = \frac{\tilde{\omega}^{N}}{N!} h_{0}(0) \left[1+\frac{q-1}{q}\frac{\Lambda+\mu N}{T}\right]^{\frac{1}{q-1}+3N},
\end{equation}
i.e., from the following equation $\partial^{2}\ln \phi(N)/\partial N^{2}=0$ for which $N=N_{0}$ is the solution. For $q>1$ the upper bound of summation $N_{0}$ follows from the Tsallis cut-off prescription~\cite{TsallCutOff}. This means that we impose the condition $1+(q-1)(\Lambda+\mu N)/qT>0$ in Eq.~(\ref{4}) and the inequality $1+(q-1)(\Lambda-p_{T} \cosh y+\mu (N+1))/qT>0$ in Eq.~(\ref{1}).

It should be stressed that in the case of $q<1$ we have two possibilities to find the cut-off parameter $N_{0}$ in Eqs.~(\ref{1}) and (\ref{4}). We can find $N_{0}$ from the local minimum of the function $\ln\phi(N)$ or from the inflection point of this function. In this paper, we fix the cut-off parameter $N_{0}$ from the inflection point of the function $\ln\phi(N)$.

Note that in the Gibbs limit $q\to 1$, Eq.~(\ref{1}) recovers the Maxwell - Boltzmann transverse momentum distribution of the Boltzmann - Gibbs statistics
\begin{equation}\label{6}
  \frac{d^{2}N}{dp_{T}dy} = \frac{gV}{(2\pi)^{2}} p_{T}^{2} \cosh y  \  e^{-\frac{p_{T} \cosh y -\mu}{T}}.
\end{equation}

The ultrarelativistic distribution function~(\ref{1}) of the Tsallis statistics given in the rapidity range $y_{0}<y<y_{1}$ can be written as~\cite{Parvan16b}
\begin{eqnarray}\label{7}
  && \left. \frac{d^{2}N}{dp_{T}dy}\right|_{y_{0}}^{y_{1}} = \frac{gV}{(2\pi)^{2}} p_{T}^{2} \int\limits_{y_{0}}^{y_{1}} dy \cosh y  \
  \sum\limits_{N=0}^{N_{0}} \frac{\tilde{\omega}^{N}}{N!} h_{0}(0) \nonumber \\
   &&\;\;\;\; \left[1+\frac{q-1}{q}\frac{\Lambda-p_{T} \cosh y +\mu (N+1)}{T}\right]^{\frac{1}{q-1}+3N}.
\end{eqnarray}

The zeroth term approximation of the Tsallis statistics for $q<1$ defined in~\cite{Parvan16b} corresponds to $N_{0}=0$ in Eqs.~(\ref{1}), (\ref{4}) and (\ref{7}). In this case, the norm function $\Lambda=0$ and the transverse momentum distribution (\ref{1}) is rewritten as
\begin{equation}\label{8}
  \frac{d^{2}N}{dp_{T}dy} = \frac{gV p_{T}^{2} \cosh y}{(2\pi)^{2}}  \left[1-\frac{q-1}{q}\frac{p_{T} \cosh y -\mu}{T}\right]^{\frac{1}{q-1}}.
\end{equation}
In the Gibbs limit $q\to 1$, Eq.~(\ref{8}) resembles the Maxwell - Boltzmann transverse momentum distribution of the Boltzmann - Gibbs statistics (\ref{6}). Note that the finite energy for the Tsallis statistics and the Tsallis statistics in the zeroth term approximation leads to the constraint that $q > 3/4$ (see Ref.~\cite{Parvan16b}).

\begin{figure}
\includegraphics[width=0.48\textwidth]{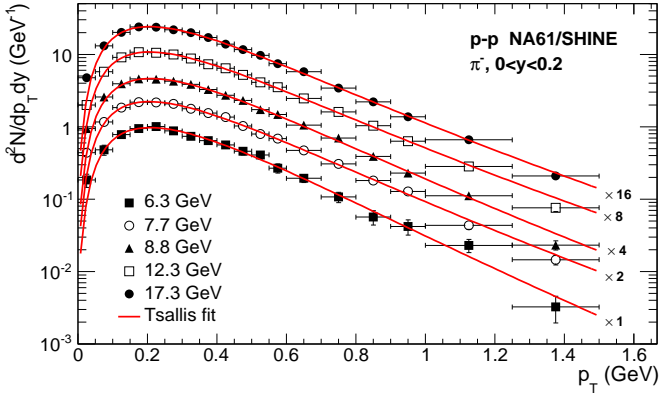}
\caption{(Color online) Transverse momentum distributions of $\pi^{-}$ pions produced in $pp$ collisions as obtained by the NA61/SHINE Collaboration~\cite{NA61} at $\sqrt{s}=6.3,7.7,8.8,12.3$ and $17.3$ GeV in the rapidity range $0<y<0.2$. The solid curves are the fits of the data to the Tsallis ultrarelativistic distribution (\ref{7}). The numbers at lines denote the scaling factor.} \label{fig3}
\end{figure}

It should be stressed that the transverse momentum distribution for the Maxwell-Boltzmann ultrarelativistic ideal gas of the Tsallis-factorized statistics, which was defined in Ref.~\cite{Cleymans2012}
\begin{equation}\label{10}
  \frac{d^{2}N}{dp_{T}dy} = \frac{gVp_{T}^{2} \cosh y}{(2\pi)^{2}}  \left[1+(q_{c}-1)\frac{p_{T} \cosh y -\mu}{T}\right]^{\frac{q_{c}}{1-q_{c}}},
\end{equation}
exactly coincides with the transverse momentum distribution (\ref{8}) of the Tsallis statistics in the zeroth term approximation with transformation of the parameter $q_{c}$ to $1/q$ (see Ref.~\cite{Parvan16b}). Thus, the momentum distribution for the ultrarelativistic ideal gas of the Tsallis-factorized statistics~\cite{Cleymans12a,Cleymans2012} is equivalent to the momentum distribution of the Tsallis statistics in the zeroth term approximation. Here the parameter $q$ from Ref.~\cite{Cleymans2012} was denoted as $q_{c}$. Note that the zeroth term approximation is valid only at large deviations of $q$ from unity~\cite{Parvan16b}. The parameter $q_{c}$ for the ultrarelativistic ideal gas of the Tsallis-factorized statistics is restricted by the condition $q_{c} < 4/3$ (see Refs.~\cite{Parvan16a,Parvan16b}).

The ultrarelativistic distribution~(\ref{10}) of the Tsallis- factorized statistics given in the rapidity range $y_{0}<y<y_{1}$ can be rewritten as
\begin{eqnarray}\label{9}
 \left. \frac{d^{2}N}{dp_{T}dy} \right|_{y_{0}}^{y_{1}} &=& \frac{gV p_{T}^{2}}{(2\pi)^{2}} \int\limits_{y_{0}}^{y_{1}} dy  \cosh y \nonumber \\
  &\times& \left[1+(q_{c}-1)\frac{p_{T} \cosh y -\mu}{T}\right]^{\frac{q_{c}}{1-q_{c}}}.
\end{eqnarray}

\begin{figure}
\includegraphics[width=0.48\textwidth]{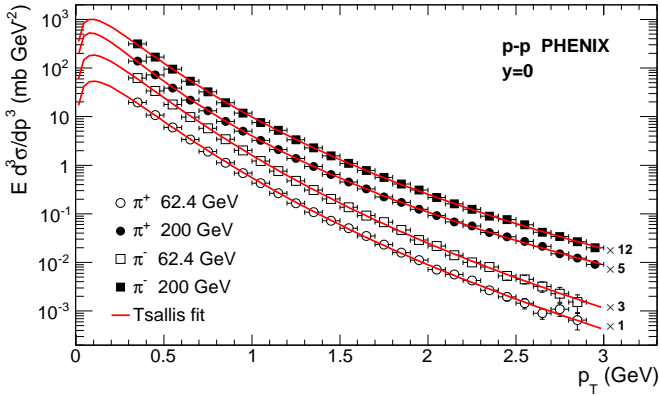}
\caption{(Color online) Transverse momentum distributions of $\pi^{-}$ and $\pi^{+}$ pions produced in $pp$ collisions as obtained by the PHENIX Collaboration~\cite{PHENIX} at $\sqrt{s}=200$ and $62.4$ GeV at midrapidity. The solid curves are the fits of the data to the Tsallis ultrarelativistic distribution (\ref{11}) at rapidity $y=0$. The numbers at lines denote the scaling factor. } \label{fig4}
\end{figure}

\begin{figure}
\includegraphics[width=0.48\textwidth]{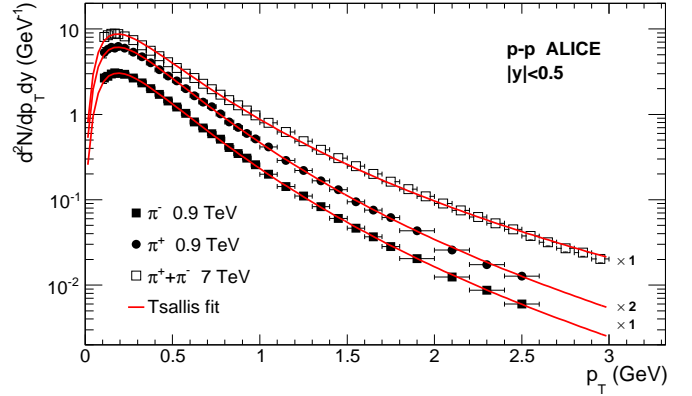}
\caption{(Color online) Transverse momentum distributions of $\pi^{-}$, $\pi^{+}$ and $\pi^{+}+\pi^{-}$ pions produced in $pp$ collisions as obtained by the ALICE Collaboration at $\sqrt{s}=0.9$ TeV~\cite{ALICE09} and $\sqrt{s}=7$ TeV~\cite{ALICE7} in the rapidity interval $|y|<0.5$. The solid curves are the fits of the data to the Tsallis ultrarelativistic distribution (\ref{7}).The numbers at lines denote the scaling factor. } \label{fig5}
\end{figure}

\begin{figure}
\includegraphics[width=0.48\textwidth]{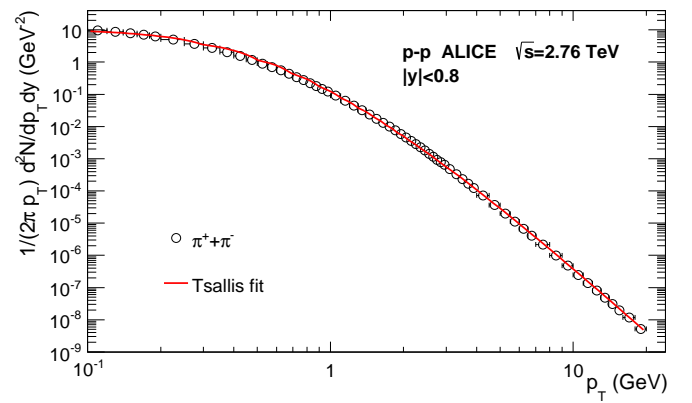}
\caption{(Color online) Transverse momentum distribution of $\pi^{+}+\pi^{-}$ pions produced in $pp$ collisions as obtained by the ALICE Collaboration~\cite{ALICE2.76} at $\sqrt{s}=2.76$ TeV in the rapidity interval $|y|<0.8$. The solid curves are the fits of the data to the Tsallis ultrarelativistic distribution (\ref{7}) divided by the geometrical factor $2\pi p_T$. } \label{fig6}
\end{figure}

Figure~\ref{fig3} represents the transverse momentum distribution of $\pi^{-}$ pions produced in the $pp$ collisions as obtained by the NA61/SHINE Collaboration~\cite{NA61} at $\sqrt{s}=6.3,7.7,8.8,12.3$ GeV and $17.3$ GeV in the rapidity interval $0<y<0.2$. The symbols represent the experimental data. The solid curves are the fits of the experimental data to the distribution function of the Tsallis statistics (\ref{7}). The values of the parameters of the distribution function (\ref{7}) of the Tsallis statistics are summarized in Table~\ref{t1} and the values of the parameters of the distribution function (\ref{9}) of the Tsallis-factorized statistics are given in Table~\ref{t2}. The Tsallis and the Tsallis-factorized curves practically coincide and give a good description of the experimental data. The transverse momentum distribution of $\pi^{-}$ pions at NA61/SHINE energies has the power law form; however, the values of the parameter $q$ are close to unity. The temperature $T$ of the massless pions for the Tsallis-factorized statistics is high and is approximately $T\sim 100$ MeV. However, the temperature $T$ of the Tsallis statistics is lower than that of the Tsallis-factorized statistics and is approximately $T\sim 80$ MeV. The values of the radius $R$ are large in comparison with the geometrical sizes of the system composed of two protons. However, the radius $R$ of the Tsallis-factorized statistics practically coincides with the radius $R$ of the Tsallis statistics and is approximately $R\sim 4.5$ fm. At the energies of the NA61/SHINE Collaboration the values of the parameter $q=1/q_{c}$ of the Tsallis-factorized statistics differ from the values of the parameter $q$ of the Tsallis statistics. Note that the experimental data of the NA61/SHINE Collaboration were fitted to the transverse momentum distribution of the Tsallis-factorized statistics for massive particles in Refs.~\cite{Rybczynski14,Parvan16a}. For the massless particles the temperature is slightly higher than the temperature of the massive particles. The radius $R$ of the system for massless particles is smaller than the radius for the massive particles and the parameter $q_{c}$ is almost the same (see Table~\ref{t2} of the present paper and Table~1 in Ref.~\cite{Parvan16a}).

Figure~\ref{fig4} represents the transverse momentum distributions of $\pi^{-}$ and $\pi^{+}$ pions produced in the proton-proton collisions as obtained by the PHENIX Collaboration~\cite{PHENIX} at $\sqrt{s}=200$ and $62.4$ GeV at midrapidity. The symbols represent the experimental data of the PHENIX Collaboration. The solid curves are the fits of the experimental data to the distribution function of the Tsallis statistics (\ref{1}) divided by  the geometrical factor $2\pi p_T$:
\begin{eqnarray}\label{11}
 &&\frac{1}{2\pi p_{T}} \frac{d^{2}N}{dp_{T}dy} = \frac{gV p_{T}\cosh y}{(2\pi)^{3}} \sum\limits_{N=0}^{N_{0}} \frac{\tilde{\omega}^{N}}{N!} h_{0}(0) \nonumber \\
   &&\;\;\;\; \left[1+\frac{q-1}{q}\frac{\Lambda-p_{T} \cosh y +\mu (N+1)}{T}\right]^{\frac{1}{q-1}+3N}. \;\;\;\;\;\;
\end{eqnarray}
The values of the parameters of the distribution function of the Tsallis statistics (\ref{11}) are given in Table~\ref{t1} and the values of the parameters for the distribution function of the Tsallis-factorized statistics (\ref{10}) divided by  the geometrical factor $2\pi p_T$ are summarized in Table~\ref{t2}. The experimental data for the transverse momentum distributions of $\pi^{-}$ and $\pi^{+}$ pions are well described by both the Tsallis statistics and the Tsallis-factorized statistics. At PHENIX energies the values of all three parameters of the transverse momentum distribution of the Tsallis-factorized statistics practically coincide with the values of the parameters of the transverse momentum distribution of the Tsallis statistics. Thus, at PHENIX energies the Tsallis-factorized statistics for massless particles satisfactorily approximates the Tsallis statistics and can be applied to describe the experimental transverse momentum distributions. Note that the experimental data of the PHENIX Collaboration for $\pi^{-}$ pions were fitted to the transverse momentum distribution of the Tsallis-factorized statistics for massive particles in Ref.~\cite{Parvan16a}. For massless particles the temperature is slightly higher than the temperature of massive particles. The radius $R$ of the system for massless particles is smaller than the radius for massive particles and the parameter $q_{c}$ is almost the same (see Table~\ref{t2} of the present paper and Table~1 in Ref.~\cite{Parvan16a}).

\begin{figure}
\includegraphics[width=0.48\textwidth]{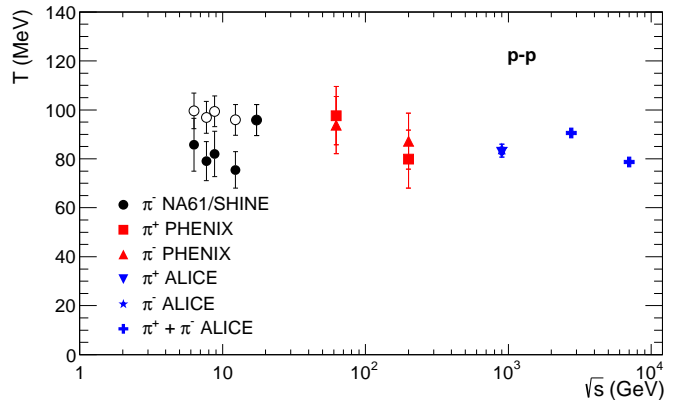}
\caption{(Color online) Energy dependence of the temperature $T$ for the Tsallis ultrarelativistic distribution. The solid points are the results of the fit by the distribution of the Tsallis statistics for the $\pi^{-}$, $\pi^{+}$ and $\pi^{+}+\pi^{-}$ pions produced in $pp$ collisions as obtained by the NA61/SHINE~\cite{NA61}, PHENIX~\cite{PHENIX} and ALICE~\cite{ALICE09,ALICE7,ALICE2.76} Collaborations. The open symbols are the results of the fit by the distribution function of the Tsallis-factorized statistics for the same data.} \label{fig7}
\end{figure}

\begin{figure}
\includegraphics[width=0.48\textwidth]{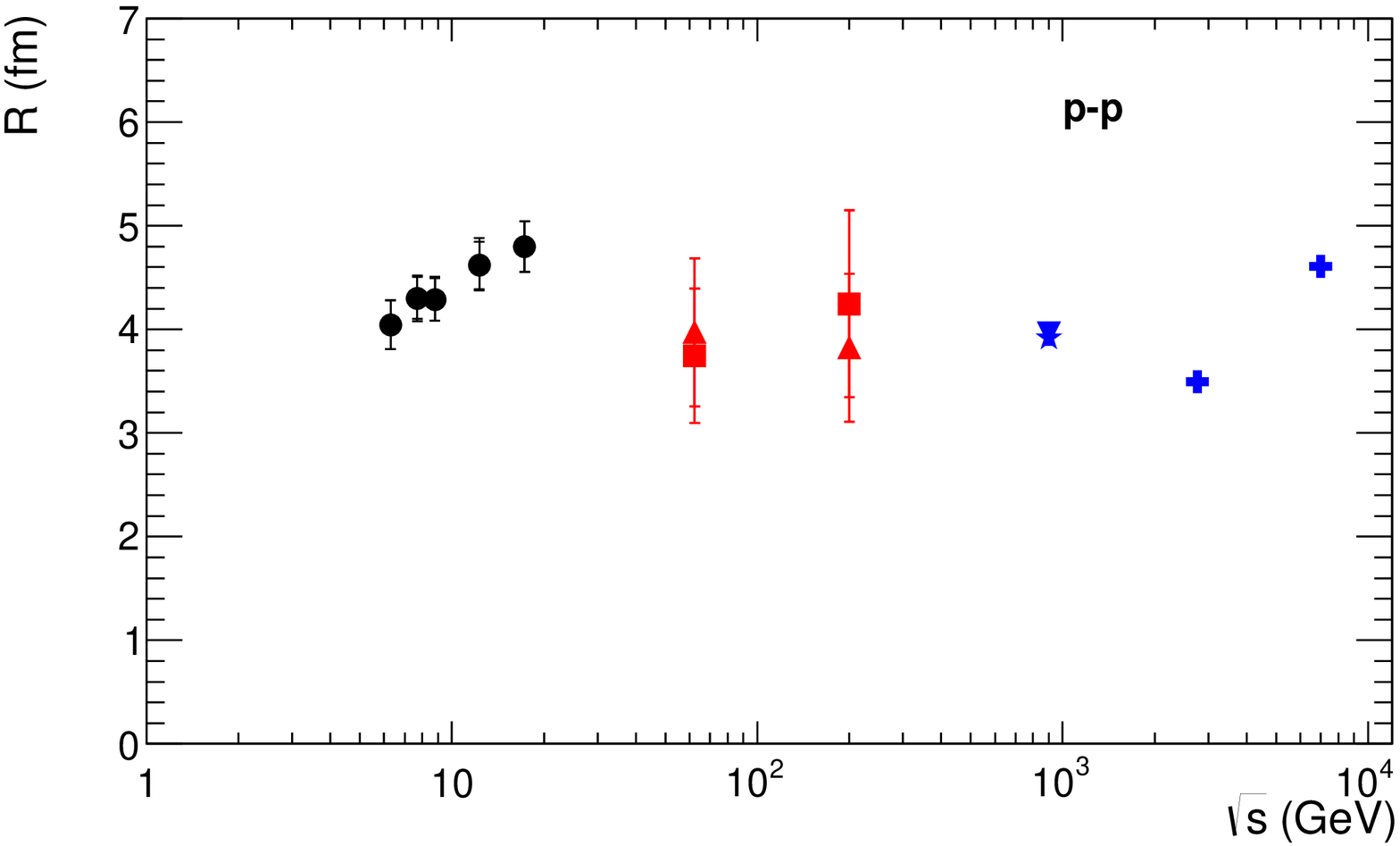}
\caption{(Color online) Energy dependence of the radius $R$ for the Tsallis ultrarelativistic distribution. The notations are the same as in Fig.~\ref{fig7}. } \label{fig8}
\end{figure}

\begin{figure}
\includegraphics[width=0.48\textwidth]{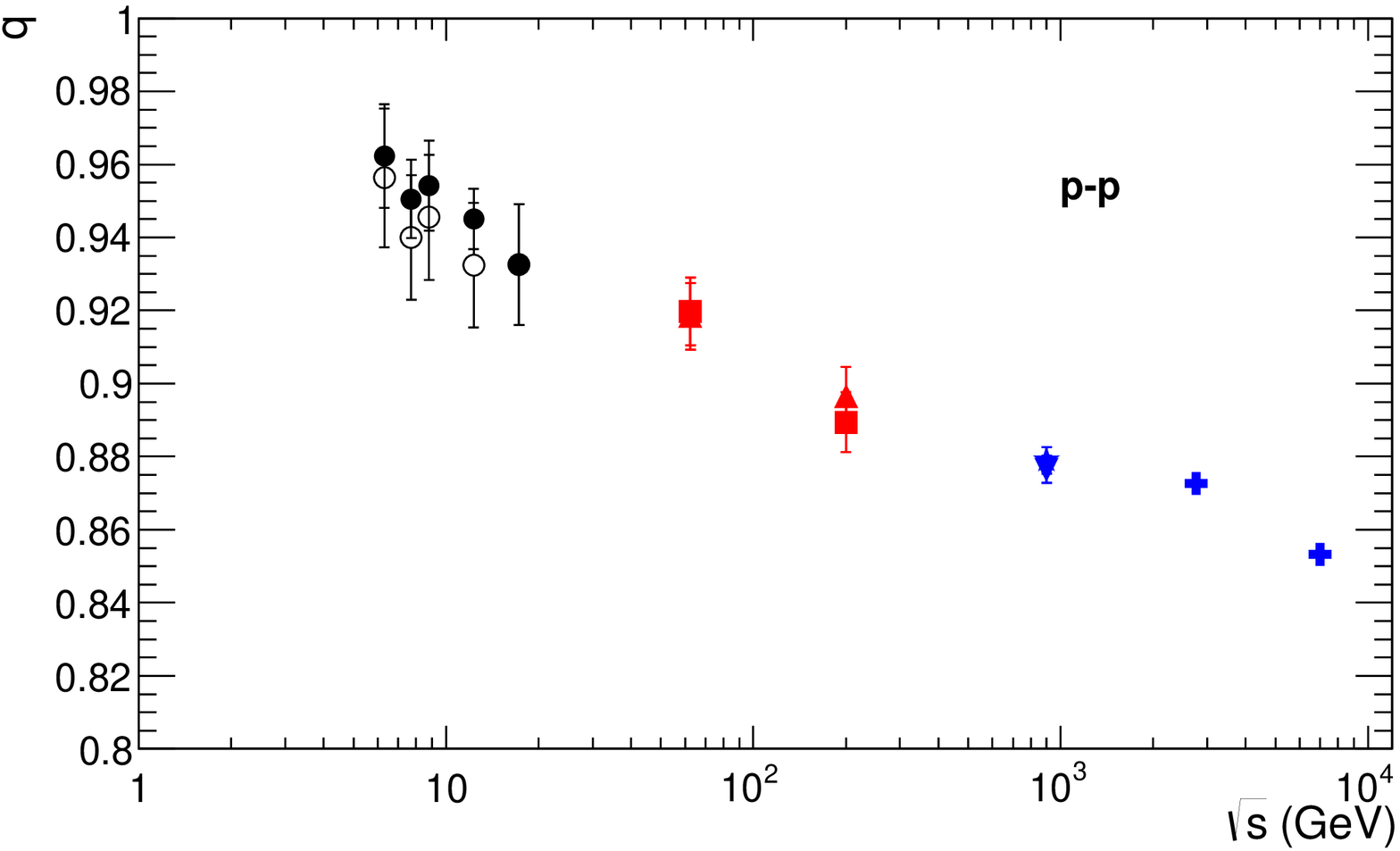}
\caption{(Color online) Energy dependence of the parameter $q$ for the Tsallis ultrarelativistic distribution. The notations are the same as in Fig.~\ref{fig7}. } \label{fig9}
\end{figure}

\begin{table*}
\begin{tabular}{rrrcccc}
 \hline
 \hline
 Collaboration  & Type & $\sqrt{s}$, GeV  &$\qquad$ $T$, MeV$ \qquad$ & $\qquad$ $\qquad$ $R$, fm$\qquad$ & $\qquad$ $\qquad$ $q$ $\qquad$ &$\qquad$$\chi^{2}/ndf$ $\qquad$ \\
 \hline
 NA61/SHINE     & $\pi^{-}$ & 6.3       & 85.78$\pm$10.79  & 4.047$\pm$0.235   & 0.9623$\pm$0.0142 & 2.821/15  \\
 NA61/SHINE     & $\pi^{-}$ & 7.7       & 79.05$\pm$8.01   & 4.304$\pm$0.204   & 0.9505$\pm$0.0107 & 1.472/15  \\
 NA61/SHINE     & $\pi^{-}$ & 8.8       & 82.01$\pm$9.28   & 4.294$\pm$0.212   & 0.9542$\pm$0.0123 & 1.821/15  \\
 NA61/SHINE     & $\pi^{-}$ & 12.3      & 75.47$\pm$7.41   & 4.627$\pm$0.253   & 0.9451$\pm$0.0083 & 1.152/15  \\
 NA61/SHINE     & $\pi^{-}$ & 17.3      & 95.83$\pm$6.38   & 4.798$\pm$0.246   & 0.9326$\pm$0.0166 & 0.865/15  \\
 PHENIX         & $\pi^{+}$ & 62.4      & 97.62$\pm$11.92  & 3.744$\pm$0.648   & 0.9197$\pm$0.0093 & 1.654/23  \\
 PHENIX         & $\pi^{-}$ & 62.4      & 93.76$\pm$11.69  & 3.971$\pm$0.716   & 0.9184$\pm$0.0091 & 0.878/23  \\
 PHENIX         & $\pi^{+}$ & 200.0     & 79.89$\pm$11.80  & 4.247$\pm$0.899   & 0.8894$\pm$0.0082 & 0.987/24  \\
 PHENIX         & $\pi^{-}$ & 200.0     & 87.20$\pm$11.48  & 3.823$\pm$0.714   & 0.8965$\pm$0.0081 & 0.691/24  \\
 ALICE          & $\pi^{+}$ & 900.0     & 82.72$\pm$2.01   & 3.965$\pm$0.069   & 0.8766$\pm$0.0037 & 3.609/30  \\
 ALICE          & $\pi^{-}$ & 900.0     & 83.92$\pm$2.02   & 3.918$\pm$0.068   & 0.8790$\pm$0.0036 & 1.610/30  \\
 ALICE          & $\pi^{+}+\pi^{-}$ & 2760.0 & 90.61$\pm$1.45  & 3.496$\pm$0.057 & 0.8726$\pm$0.0012 & 12.18/60  \\
 ALICE          & $\pi^{+}+\pi^{-}$ & 7000.0 & 78.75$\pm$1.86  & 4.606$\pm$0.093 & 0.8533$\pm$0.0024 & 9.775/38  \\
\hline
\hline
\end{tabular}
\caption{Parameters of the Tsallis statistics fit for the pions produced in $pp$ collisions at different energies.}
\label{t1}
\end{table*}

\begin{table*}
\begin{tabular}{rrrccccc}
 \hline
 \hline
 Collaboration  & Type & $\sqrt{s}$, GeV  &$\quad$ $T$, MeV$ \quad$ & $\quad$ $\quad$ $R$, fm$\quad$ & $\quad$ $\quad$ $q=1/q_{c}$ $\quad$ & $q_{c}$ &$\quad$$\chi^{2}/ndf$ $\quad$ \\
 \hline
 NA61/SHINE     & $\pi^{-}$ & 6.3       & 99.59$\pm$7.32   & 4.045$\pm$0.234   & 0.9563$\pm$0.0190 & 1.0457$\pm$0.0208 & 2.825/15  \\
 NA61/SHINE     & $\pi^{-}$ & 7.7       & 96.93$\pm$6.49   & 4.300$\pm$0.222   & 0.9400$\pm$0.0171 & 1.0638$\pm$0.0194 & 1.481/15  \\
 NA61/SHINE     & $\pi^{-}$ & 8.8       & 99.37$\pm$6.29   & 4.290$\pm$0.204   & 0.9455$\pm$0.0172 & 1.0576$\pm$0.0193 & 1.838/15  \\
 NA61/SHINE     & $\pi^{-}$ & 12.3      & 95.92$\pm$6.29   & 4.619$\pm$0.228   & 0.9324$\pm$0.0170 & 1.0725$\pm$0.0196 & 1.175/15  \\
 NA61/SHINE     & $\pi^{-}$ & 17.3      & 95.83$\pm$6.38   & 4.798$\pm$0.246   & 0.9326$\pm$0.0166 & 1.0722$\pm$0.0191 & 0.865/15  \\
 PHENIX         & $\pi^{+}$ & 62.4      & 97.62$\pm$11.92  & 3.744$\pm$0.648   & 0.9197$\pm$0.0093 & 1.0874$\pm$0.0110 & 1.654/23  \\
 PHENIX         & $\pi^{-}$ & 62.4      & 93.76$\pm$11.69  & 3.971$\pm$0.715   & 0.9184$\pm$0.0091 & 1.0888$\pm$0.0108 & 0.878/23  \\
 PHENIX         & $\pi^{+}$ & 200.0     & 79.89$\pm$11.81  & 4.247$\pm$0.899   & 0.8894$\pm$0.0082 & 1.1244$\pm$0.0104 & 0.987/24  \\
 PHENIX         & $\pi^{-}$ & 200.0     & 87.20$\pm$11.49  & 3.823$\pm$0.714   & 0.8965$\pm$0.0081 & 1.1155$\pm$0.0101 & 0.691/24  \\
 ALICE          & $\pi^{+}$ & 900.0     & 82.72$\pm$2.01   & 3.965$\pm$0.069   & 0.8766$\pm$0.0037 & 1.1408$\pm$0.0048 & 3.609/30  \\
 ALICE          & $\pi^{-}$ & 900.0     & 83.92$\pm$2.02   & 3.918$\pm$0.068   & 0.8790$\pm$0.0036 & 1.1376$\pm$0.0047 & 1.610/30  \\
 ALICE          & $\pi^{+}+\pi^{-}$ & 2760.0 & 90.61$\pm$1.45  & 3.496$\pm$0.057 & 0.8726$\pm$0.0012 & 1.1460$\pm$0.0016 & 12.18/60  \\
 ALICE          & $\pi^{+}+\pi^{-}$ & 7000.0 & 78.75$\pm$1.86  & 4.606$\pm$0.093 & 0.8533$\pm$0.0024 & 1.1719$\pm$0.0032 & 9.775/38  \\
\hline
\hline
\end{tabular}
\caption{Parameters of the fit by the distribution function of the Tsallis-factorized statistics for the charged pions produced in $pp$ collisions at different energies.}
\label{t2}
\end{table*}

Figures~\ref{fig5} and~\ref{fig6} represent the transverse momentum distributions of $\pi^{-}$, $\pi^{+}$ and $\pi^{+}+\pi^{-}$ pions produced in the $pp$ collisions as obtained by the ALICE Collaboration at $\sqrt{s}=0.9$ TeV~\cite{ALICE09} and $7$ TeV~\cite{ALICE7} in the rapidity interval $|y|<0.5$, and at $\sqrt{s}=2.76$ TeV~\cite{ALICE2.76} in the rapidity interval $|y|<0.8$. The symbols represent the experimental data. In Fig.~\ref{fig5}, the solid curves are the fits of the experimental data to the distribution function of the Tsallis statistics (\ref{7}). In Fig.~\ref{fig6}, the solid curve is the fit of the experimental data to the distribution function of the Tsallis statistics (\ref{7}) divided by the geometrical factor $2\pi p_T$. The values of the parameters of the distribution function of the Tsallis statistics (\ref{7}) and of the function (\ref{7}) divided by the geometrical factor $2\pi p_T$  are given in Table~\ref{t1}. The values of the parameters for the distribution function of the Tsallis-factorized statistics (\ref{9}) and of the function (\ref{9}) divided by the geometrical factor $2\pi p_T$ are given in Table~\ref{t2}. The experimental data of the ALICE Collaboration are well described by both the Tsallis statistics and the Tsallis-factorized statistics in the whole $p_{T}$ region. At energies of the ALICE Collaboration from $\sqrt{s}=0.9$ TeV to $7$ TeV the values of the parameters of the transverse momentum distribution of the Tsallis-factorized statistics coincide with the values of the parameters of the transverse momentum distribution of the Tsallis statistics. Thus, at ALICE energies the Tsallis-factorized statistics for massless particles well approximates the Tsallis statistics and can be applied to describe the experimental transverse momentum distributions. Note that the experimental data for $\pi^{-}$ pions of the ALICE Collaboration at $\sqrt{s}=0.9$ TeV  were fitted to the transverse momentum distribution of the Tsallis-factorized statistics for massive particles in Ref.~\cite{Parvan16a}. For massless particles the temperature is higher than the temperature of massive particles. The radius $R$ of the system in the case of massless particles is smaller than the radius for massive particles. And the parameter $q_{c}$ for massless particles is slightly smaller than the parameter $q_{c}$ of massive ones (see Table~\ref{t2} of the present paper and Table~1 in Ref.~\cite{Parvan16a}).

\begin{figure*}
\includegraphics[width=0.99\textwidth]{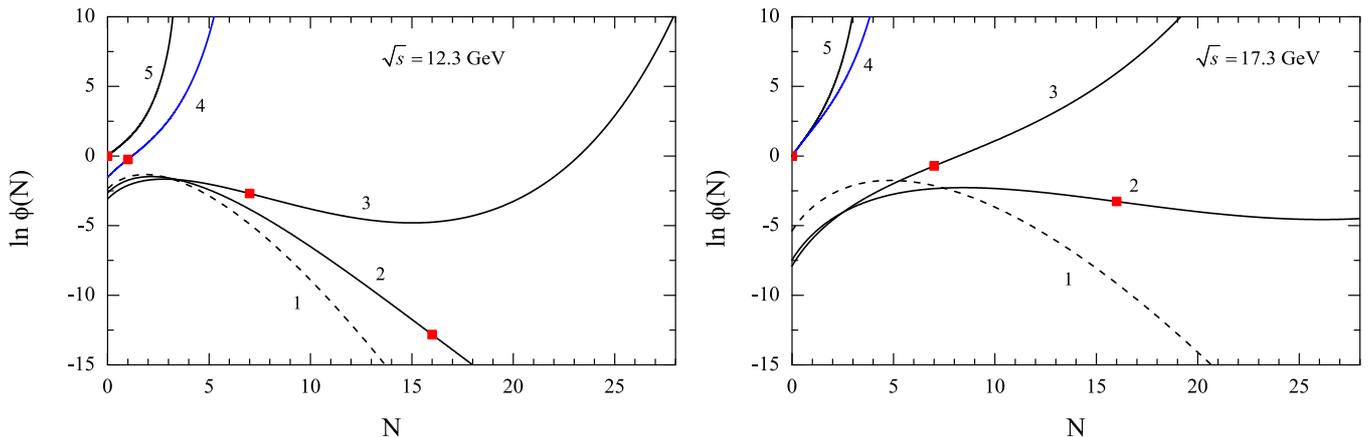}
\caption{(Color online) The cut-off parameter $N_{0}$ for the Tsallis statistics at two different energies. The behavior of $\ln\phi(N)$ as a function of $N$ at $\sqrt{s}=12.3$ GeV (left panel) and $\sqrt{s}=17.3$ GeV (right panel), and different values of the parameter $q$. The lines $1,2,3$ and $5$ are calculations for $q=1,0.995,0.99$ and $0.9$, respectively. The line $4$ corresponds to $q=0.9451$ (left panel) and $q=0.9326$ (right panel). Symbols represent the cut-off parameter $N_{0}$. The values of the parameters $T$ and $R$ are given in Table~\ref{t1}.    } \label{fig10}
\end{figure*}

The energy dependence of the parameters of both the Tsallis distribution and the Tsallis-factorized distribution for $\pi^{-}$, $\pi^{+}$ and $\pi^{+}+\pi^{-}$ pions produced in $pp$ collisions is shown in Figs.~\ref{fig7}, \ref{fig8} and \ref{fig9}. The values of the parameters of the Tsallis distribution are compared with the values of the parameters of the Tsallis-factorized statistics. The solid points are the results of the fit for the Tsallis statistics. The open points are the results of the fit for the Tsallis-factorized statistics. Figure~\ref{fig7} represents the energy dependence of the temperature $T$ for the Tsallis ultrarelativistic distribution. It is clearly seen that the temperature $T$ of the Tsallis-factorized statistics differs from the temperature of the Tsallis statistics only at energies of the NA61/SHINE Collaboration. The difference between the values of these two temperatures disappears with $\sqrt{s}$. The temperature $T$ of the massless pions slowly decreases with energy of collision. The values of the radius $R$ of the Tsallis-factorized statistics practically coincide with the values of the radius $R$ for the Tsallis statistics. See Fig.~\ref{fig8}. The radius $R$ for the distribution of the massless pions is independent of the energy of collision. The energy dependence of the parameter $q$ is represented in Fig.~\ref{fig9}. The transverse momentum distribution for the massless pions has a power-law form and increasingly deviates from the exponential function with collision energy, i.e. the parameter $q$ is not equal to unity and decreases significantly with $\sqrt{s}$. The evident difference between the Tsallis-factorized statistics and the Tsallis statistics is clearly seen only at low energies of the NA61/SHINE Collaboration. At higher energies of the PHENIX and ALICE Collaborations this difference completely disappears. Thus, we can conclude that in the expected NICA and FAIR energy range, which is close to the energy range of the NA61/SHINE Collaboration, the transverse momentum distributions of the Tsallis statistics should be used instead of the transverse momentum distributions of the Tsallis-factorized statistics given in~\cite{Cleymans12a,Cleymans2012}.

It should be stressed that the fitted temperature of the Tsallis statistics for the NA61/SHINE data has an unphysical jump at $\sqrt{s}=17.3$ GeV. See Fig.~\ref{fig7} and Table~\ref{t1}. Such a large jump in temperature may be entirely explained by the mathematical properties of the Tsallis statistics. The sums in Eqs.~(\ref{1}) and (\ref{4}) for the Tsallis statistics are divergent for $q<1$. We regularize them by the cut-off procedure described above excluding the unphysical terms with $N>N_{0}$. The number of physical terms $N_{0}+1$ in these sums depends on the values of the parameters $(T,V,\mu,q)$. See, for example, Fig.~\ref{fig10}. This number of terms in the sums~(\ref{1}) and (\ref{4}) decreases discretely up to 1 with increasing the deviation of the parameter $q$ from unity. One remaining term in these sums at small values of $q$ corresponds to the zeroth term approximation of the Tsallis statistics, the momentum distribution function of which is equivalent to the distribution function of the Tsallis-factorized statistics. Figure~\ref{fig10} clearly shows that the sums~(\ref{1}) and (\ref{4}) of the Tsallis statistics contain two physical terms $(N_{0}=1)$ at $\sqrt{s}=12.3$ GeV and one physical term $(N_{0}=0)$ at $\sqrt{s}=17.3$ GeV. Thus, the change in the number of terms in these sums from two to one for $\sqrt{s}=12.3$ and $17.3$ GeV leads to a jump in the fitted temperature $T$ of the Tsallis statistics. This is not a physical result but represents a mathematical problem of the original Tsallis statistics\cite{Tsal88,Tsal98} in describing the transverse momentum distributions at high energies because the divergent series of the Tsallis statistics are truncated by the cut-off parameter $N_{0}$, which depends on the variables of state of the system.

\section{Discussion and conclusions}\label{sec5}
In the present paper, the Tsallis statistics was applied to describe the experimental data on the transverse momentum distributions of hadrons. We have numerically compared the transverse momentum distribution of the Tsallis statistics with the transverse momentum distribution of the Tsallis-factorized statistics and applied them to describe the experimental data of the charged pions produced in $pp$ collisions at energies of the NA61/SHINE, PHENIX and ALICE Collaborations. The parameters of the Tsallis statistics and the Tsallis- factorized statistics were obtained. It was found that the results of the Tsallis-factorized statistics deviate from the results of the Tsallis statistics only at low energies of the NA61/ SHINE Collaboration when the values of the parameters $q$ and $q_{c}$ are close to unity. At PHENIX and ALICE energies, which are higher, and when the values of the parameters $q$ and $q_{c}$ deviate essentially from unity, the Tsallis-factorized statistics recovers the results of the Tsallis statistics. Thus, we can conclude that in contrast to the Tsallis-factorized statistics the Tsallis statistics will be important to the description of the experimental data on the transverse momentum spectra of hadrons at the future NICA and FAIR colliders.

\begin{acknowledgement}
This work was supported in part by the joint research project and grant of JINR and IFIN-HH (protocol N~4543). I am indebted to D.-V.~Anghel, J.~Cleymans, G.I.~Lykasov, A.S.~Sorin and O.V.~Teryaev for stimulating discussions.
\end{acknowledgement}

\end{document}